\documentclass[useAMS,usenatbib]{mn2e}
\usepackage{graphicx}
\title{High Dispersion Absorption-line Spectroscopy of AE Aqr}

\author[Echevarr\'{\i}a et al.]
  {J.~Echevarr\'{\i}a,$^1$\thanks{jer@astroscu.unam.mx}
  Robert Connon Smith,$^3$
  R.~Costero,$^1$
  S. Zharikov,$^2$
  R. Michel$^2$\\
    $^1$Instituto de Astronom{\'\i}a, Universidad Nacional Aut\'onoma de
  M\'exico, Apartado Postal 70-264, M\'exico, D.F., 04510, M\'exico\\
  $^2$Instituto de Astronom{\'\i}a, Universidad Nacional Aut\'onoma de
  M\'exico, Apartado Postal 877, Ensenada, Baja California, 22800, M\'exico\\
  $^3$Department of Physics \&\ Astronomy, University of Sussex,
  Falmer, Brighton, Sussex BN1 9QH, UK\\
}
\date{Accepted 2008 xxxxx.
      Received \today}

\pagerange{\pageref{firstpage}--\pageref{lastpage}}
\pubyear{2007}

\begin{document}

\maketitle

\label{firstpage}

\begin{abstract}

High-dispersion time-resolved spectroscopy of the unique magnetic
cataclysmic variable AE~Aqr is presented. A radial velocity analysis
of the absorption lines yields $K_2 = 168.7 \pm 1 $\,km\,s$^{-1}$.
Substantial deviations of the radial velocity curve from a sinusoid
are interpreted in terms of intensity variations over the secondary
star's surface. A complex rotational velocity curve as a function of
orbital phase is detected which has a modulation frequency of twice
the orbital frequency, leading to an estimate of the binary
inclination angle that is close to $70^\circ$. The minimum and
maximum rotational velocities are used to indirectly derive a mass
ratio of $q= 0.6$ and a radial velocity semi-amplitude of the white
dwarf of $K_1 = 101 \pm 3 \,$\,km\,s$^{-1}$. We present an
atmospheric temperature indicator, based on the absorption line
ratio of Fe~I and Cr~I lines, whose variation indicates that the
secondary star varies from K0 to K4 as a function of orbital phase.
The ephemeris of the system has been revised, using more than one
thousand radial velocity measurements, published over nearly five
decades. From the derived radial velocity semi-amplitudes and the
estimated inclination angle, we calculate that the masses of the
stars are $M_1 = 0.63 \pm 0.05 \, $M$_{\odot}$;  $ M_2 = 0.37 \pm
0.04 \, $M$_{\odot} $, and their separation is $ a = 2.33 \pm 0.02
\, $R$_{\odot}$. Our analysis indicates the presence of a late-type
star whose radius is larger, by a factor of nearly two, than the
radius of a normal main sequence star of its mass. Finally we
discuss the possibility that the measured variations in the
rotational velocity, temperature, and spectral type of the secondary
star as functions of orbital phase may, like the radial velocity
variations, be attributable to regions of enhanced absorption on the
star's surface.

\end{abstract}


\section{Introduction}  \label{intro}

Cataclysmic variables are semi-detached binary systems in which a
red dwarf secondary loses material through the inner Lagrangian
point into an accretion disc, ring or accretion column around the
white dwarf primary. The secondary stars in cataclysmic variables
are challenging targets to study because their spectrum is heavily
veiled by the strong blue continuum of the accretion disc. Since the
donor star is an important piece in the understanding of these
interacting binaries, it is paramount to make detailed and high
quality spectroscopic studies of their secondary stars.

AE Aquarii is an 11-12 mag cataclysmic variable, which was
discovered in the optical by Zinner (1938). It is located at about
100 pc (Friedjung~1997) and was first associated with the DQ~Her or
{\it Intermediate Polar} stars by Patterson~(1979). Its binary
nature was discovered by Joy~(1954), who found this system to be
consistent with a small hot star and a late-type companion of
spectral type dK0, with an orbital period of about 16.84 hr. A
corrected period, of around 9.88 hr, was obtained later by Walker
(1965). Since AE~Aqr is not an eclipsing system, its inclination
angle is not well determined, but it is thought to be $ i \approx
60^\circ \pm 10^\circ$, while its mass ratio is in the range $ 0.6
\leq q \leq 0.84$ (Patterson 1979; Chincarini \& Walker 1981;
Robinson, Shafter \& Balachandran 1991; Reinsch \& Beuermann 1994;
Welsh, Horne \& Gomer 1995; Casares et al. 1996; Watson, Dhillon \&
Shahbaz 2006; this paper). The absorption lines from the secondary
star are very strong. The light curve of AE~Aqr exhibits large
flares and coherent oscillations of about 16 and 33~s in the optical
and X-ray (Patterson 1979). It also exhibits radio and millimetre
synchrotron emission (e.g. Bookbinder \& Lamb 1987; Bastian, Dulk \&
Chanmugam~1988), as well as TeV $\gamma$-rays (Bowden et al. 1992;
Meintjes et al. 1992). The Balmer emission lines vary both in
strength and shape and they may not be good tracers of the orbital
motion of the white dwarf. This has led to the proposal of the
magnetic propeller model (Wynn, King \& Horne 1997).

The study of the physical conditions in the intermediate polar
AE~Aqr is important because the system shows some remarkable and
unique features. Understanding the behaviour of this binary may help
us to understand, for example, the role that mass transfer has in
this kind of binary (Watson et al. 2006). Obtaining accurate masses
in this system is also an important goal. Because the secondary
spectrum is clearly visible, it is possible to determine a reliable
value for $K_2$, as we do below. In AE~Aqr, an unstable accretion
disc is present, which makes direct observations of the radial
velocity semi-amplitude of the accretion disc difficult and,
consequently, the direct determination of the real $K_1$ value is
unreliable. One indirect method of determining $K_1$ is to use the
orbital variations in the spin pulse of the rapidly rotating white
dwarf (Robinson et al. 1991). However, given that the secondary star
shows a clearly measurable rotational velocity, an alternative way
of determining the mass ratio is by combining the $K_2$ value with
the rotational velocity of the late-type star (Horne, Wade \& Szkody
1986). It is then possible to deduce the value of $K_1$, and use it,
along with our measured value of $K_2$ and our estimate of the
inclination, to calculate the individual stellar masses (cf. Section
\ref{param}).

Detailed radial velocity studies of the secondary star were first
made by Chincarini \& Walker (1981) and later by Robinson et al.
(1991), Reinsch \& Beuermann (1994), Welsh et al. (1995), Casares et
al. (1996) and Watson et al. (2006). We present here
observations of AE~Aqr,  made with the UCL echelle spectrograph
(UCLES) at the Anglo Australian Telescope and with the Echelle
spectrograph at the San Pedro M\'artir Observatory in M\'exico.
In this first paper, we concentrate on the radial velocity
analysis of the absorption lines and compare our results with
previous studies. In a later paper, after obtaining further
observations, we will deal with the complex emission lines, whose
analysis imposes significant challenges; the lines sometimes appear
to come from fragmented material orbiting the white dwarf, while at
other times they show the presence of a full accretion disc.

\section{Observations}  \label{obs}

\begin{table*}
 \centering
   \caption{Parameters of Spectral Type Template Stars}
    \label{stand}
   \begin{minipage}{150mm}
   \begin{tabular}{llcccl}
   \hline
    Name  & Henry & Spectral &  $\rmn{V_{Rad}}$ & ($\rmn{V_{Rad}})^{4}$ & Ref%
 \footnote{ References to Spectral Types: (a)  Houk \& Smith-Moore 1988; (b) Houk 1982; (c) Wilson 1962;
 (d) Morgan \& Keenan 1973; (e) Morgan, Keenan \& Kellman 1943; (f) Johnson \& Morgan 1953;
 (g) Roman 1955; (h) Evans, Menzies \& Stoy 1957; (i) Roman 1952; (j) Hoffleit 1964.
 References to Radial Velocities: (1) Wilson 1953; (2) Evans 1967; (3) Flynn \& Freeman 1993;
 (4) This paper. Note: The spectral type in the last column, given in
parentheses, refers to the value derived in this paper.}\\

       & Draper &  Type   &    \multicolumn{2}{c}{($\rmn{km\,s^{-1}}$)}     \\
\hline
            \noalign{\smallskip}
--21 4065     & HD135534 & K2 III    &    --5.3     &  --9.7     &  a,1      \\
36 Lib       & HD138688 & K3 III    &    12.3     &   6.1     &  b,1      \\
36 Oph A     & HD155886 & K0 V      &    --0.6     &   0.4     &  h,2      \\
36 Oph B     & HD155885 & K2 V      &     0.0     &  --0.4    &  c,2 (col 3: G8 V)  \\
3 Ser        & HD135482 & K0 III    &   --34.24    &    -     &  j,1  \\
$\delta$ Oph & HD146051 & M0.5 III  &   --19.9     & --21.6    &  d,1      \\
$\eta$ Scl   & HD2429   & M2/M4 III &    11.2     &  12.2    &  b,1      \\
HR 1004      & HD20729  & M1 III    &    +15      &  11.7     &  a,1      \\
HR 13        & HD344    & K1 III    &     +7      &   7.2     &  j,3      \\
HR 222       & HD4628   & K2 V      &    --12.6    &  --9.8     &  e,1      \\
HR 471       & HD10142  & K0 III    &     29.7    &  29.1     &  b,2      \\
HR 487       & HD10361  & K0 V      &     19.5    &  20.8    &  h,2      \\
HR 486       & HD10360  & K0 V      &     22.5    &  22.2    &  h,2      \\
HR 54        & HD1089   & K3 III    &     --52     & --51.6    &  j,3      \\
HR 67        & HD1367   & K0 II     &     --9.4    &  --8.8     &  j,1      \\
HR 753       & HD16160  & K3 V      &     23.4    &  24.4   &  g,1      \\
$\kappa$ Oph & HD153210 & K2 IIIvar &    --55.6    & --57.2       &  i,1   \\
61 Cyg A     & HD201091 & K5 V      &    --64.3    &   -      &  f,1      \\
61 Cyg B     & HD201092 & K7 V      &    --63.5    &   -       &  f,1      \\
\hline\hline
\end{tabular}
\end{minipage}
\end{table*}

Spectroscopic observations of AE Aqr were made using the
Anglo-Australian Telescope (AAT) and the University College London
Echelle Spectrograph (UCLES) at the coud\'e focus on 1991 August 2
and 3. We obtained 102 spectra, 360~s exposure each, with a typical
signal-to-noise ratio of 10 (around  $\lambda$4500~\AA). We
used the 31.6 lines~mm$^{-1}$ grating and the Blue Thomson
1024$\times$1024 CCD with the 700~mm camera. The spectral region
$\lambda$4000~\AA~ to $\lambda$5100~\AA~ was covered. The spectral
resolution was about 5.4\,km\,s$^{-1}$. Almost two complete orbital
periods were covered. Several late-type template stars were
also observed.

Three further runs were obtained at the Observatorio Astron\'omico
Nacional at San Pedro M\'artir (SPM) using the 2.1m Telescope and
the Echelle Spectrograph. The first observations were gathered on
the nights of 1997 September 22 and 23. We obtained 18 spectra,
600~s exposure, with typical signal-to-noise ratio of 25 (around
$\lambda$5450~\AA). We used the $15 \mu$m Thomson 2048$\times$2048
detector, with the 300 lines~mm$^{-1}$ echellette grating to cover a
spectral range from $\lambda$3400~\AA~to $\lambda$7300~\AA. The
spectral resolution for this configuration was about
15.4\,km\,s$^{-1}$. Orbital phases 0.37 -- 0.55 (first day) and
0.58 -- 1.00 (second day) were covered in this run.  In the
second set of observations, made on the nights of 2000 August 17 to
19, 81 spectra of 600~s exposure were obtained. The instrumental
setup was similar to the first one, except that we covered a
spectral range from $\lambda$3700~\AA~to $\lambda$7700~\AA. Orbital
phases $0.35 - 0.90$, $0.58 - 1.36$ and $0.0 - 0.73$ were covered
on successive days with a typical signal-to-noise
ratio of 42 (around $ \lambda \, 5450$ \AA). The third run was made
on 2001 August 29, when 47 spectra were obtained with shorter
exposure times of 180~s with a typical signal-to-noise ratio of 28
(around $\lambda$5450~\AA). We used the $24 \mu$m SITe
1024$\times$1024 detector, with a configuration to cover a spectral
range from $\lambda$3900~\AA~to $\lambda$7100~\AA. The spectral
resolution for this configuration is about 24.6\,km\,s$^{-1}$. Only
two spectral type standards, 61 Cygni A and B, were observed during
the SPM runs.

The list of observed standards is shown in Table~\ref{stand}. Most
of these stars are not primary standards. Their spectral
classifications (column 3) have been taken mainly from the Bright
Star Catalogue and updated from several sources (see references in
the Table). Their radial velocities are also shown; these values
have been obtained from a variety of sources, as listed in the Table
references. We have also derived our own radial velocity
measurements (column 5), and have cross-correlated the results to
check for inconsistencies. In general there is a good agreement
among all sources and we have adopted the values given in column 4.
The published rotational velocities, not listed, all have very low
values, less than our spectral resolution of 10\,km\,s$^{-1}$, as
expected in general from isolated cool main-sequence stars.

\section{Data Reduction and Spectral Features}  \label{reduc}

The AAT and SPM spectra were treated in a similar manner. They
both have high dispersion orders separated by a cross-disperser, which
have to be first extracted into one dimensional spectra, wavelength
calibrated separately and then merged into a single array and
re-binned to a single linear dispersion. The great stability of both
UCLES and the Echelle Spectrograph allows us to interpolate with
ease the calibration lamp spectra taken during the observing nights.
The AAT spectra were reduced with STARLINK software and the SPM
observations were reduced with the NOAO/IRAF V2.11 package\footnote
{IRAF is distributed by the National Optical Astronomical
Observatories, operated by the Association of Universities for
Research in Astronomy, Inc., under cooperative agreement with the
National Science Foundation.}.

The spectra of AE Aqr (e.g. Casares et al. 1996) show
complex H$\beta$ and H$\gamma$ emission lines and a weak HeI
$\lambda 4471$, superimposed on a red continuum with strong
absorption lines, mainly from FeI, CrI, CaI, and MnI. As
discussed in the Introduction, we shall leave the emission lines to
a later paper and concentrate here on the behaviour and
interpretation of the absorption lines coming from the secondary
star.

The radial velocities were obtained for all observations using the
{\it fxc} routine from the {\it rv} package in IRAF. For the case of
the AAT data we selected for analysis the spectral region
$\lambda4200$ \AA \, to $\lambda4330$ \AA, while for the SPM data we
selected the region $ \lambda4870$ \AA  \, to $\lambda5530$ \AA. The
selected spectral regions in the AAT and SPM data are not the same.
For the SPM data we have combined the echelle orders which, due to
the instrumental setup, have the maximum fluxes, while in the case
of the AAT data, we have selected a region where the free
spectral ranges overlap; this also avoids contamination from the
H$\gamma$ line. As we will show in the analysis, the different
selection for the {\it blue} and {\it green} spectral regions
contributes a check on the validity of the results.  We made a check
for consistency for the AAT and SPM data with all the template
stars. The semi-amplitude of the secondary obtained from the
different standards shows a range of about 3\,km\,s$^{-1}$, with no
obvious correlation with spectral type. The cross-correlation peaks
obtained with the {\it fxc} routine were well fitted by Gaussian
functions in all cases.

\section{Radial Velocities and Spectroscopic Orbital Parameters}  \label{radvel}

The measured radial velocities are shown in Tables
\ref{RadVel1}~and~\ref{RadVel2} for the AAT observations with
respect to HR~222, one of the observed standards which has a K2~V
spectral type, consistent with the observed spectral type (see
section~\ref{sptype}). The radial velocities for the SPM data are
shown in Tables~\ref{RadVel3}~to~\ref{RadVel5} with respect to
61~Cyg~A.

   \begin{table}
      \caption{Radial velocities for AE Aqr for August 2, 1991}
         \label{RadVel1}
         \begin{tabular}{rrr}
            \hline
            \noalign{\smallskip}
        HJD     & Absorption    & $\sigma $   \\
        (240000+)  &    (km\,s$^{-1}$)  \\
            \noalign{\smallskip}
            \hline
            \noalign{\smallskip}
48470.9548 &   91.0710 &  6.851  \\
48470.9601 &   96.3983 &    6.736  \\
48470.9654 &   94.5199 &    6.826  \\
48470.9707 &   96.6956 &    6.836  \\
48470.9760 &   95.4723 &    7.065  \\
48470.9813 &   95.0824 &    6.284  \\
48470.9866 &   96.2528 &    6.969  \\
48470.9919 &   91.3255 &    7.177  \\
48470.9972 &   87.0076 &    7.091  \\
48471.0025 &   81.8741 &    6.925  \\
48471.0108 &   72.1152 &    7.950  \\
48471.0161 &   65.8743 &    8.984  \\
48471.0214 &   59.1111 &   10.617 \\
48471.0267 &   53.4765 &   13.676 \\
48471.0320 &   41.8962 &   10.892 \\
48471.0373 &   31.6189 &    8.460  \\
48471.0426 &   21.4795 &    7.380  \\
48471.0479 &   10.5793 &    5.600  \\
48471.0532 &   -0.8366 &    4.731  \\
48471.0585 &  -15.2109 &    4.909  \\
48471.0638 &  -28.1980 &    5.468  \\
48471.0691 &  -39.2205 &    5.440  \\
48471.0744 &  -54.6164 &    6.714  \\
48471.0797 &  -71.7725 &    6.748  \\
48471.0850 &  -89.5616 &    7.276  \\
48471.0903 &  -107.0077 &    7.192  \\
48471.0956 &  -119.2595 &    6.615  \\
48471.1009 &  -134.7471 &    6.247  \\
48471.1062 &  -147.4073 &    6.071  \\
48471.1115 &  -157.5921 &    5.689  \\
48471.1168 &  -169.2017 &    5.870  \\
48471.1221 &  -179.3889 &    5.770  \\
48471.1275 &  -188.2508 &    5.805  \\
48471.1328 &  -193.2074 &    5.509  \\
48471.1381 &  -202.0784 &    6.967  \\
48471.1434 &  -208.8203 &    6.023  \\
48471.1487 &  -213.4168 &    6.744  \\
48471.1540 &  -218.1144 &    7.078  \\
48471.1644 &  -228.3875 &   11.888 \\
            \noalign{\smallskip}
            \hline
         \end{tabular}
   \end{table}

   \begin{table}
      \caption[]{Radial velocities for AE Aqr for August 3, 1991}
         \label{RadVel2}
      \[
         \begin{tabular}{rrr}
            \hline
            \noalign{\smallskip}
        HJD     & Absorption    & $\sigma $   \\
        (240000+)  &    (km\,s$^{-1}$)  \\
            \noalign{\smallskip}
            \hline
            \noalign{\smallskip}

48471.9201 &     -126.4122 &    8.231\\
48471.9259 &     -139.3585 &    8.008  \\
48471.9312 &     -155.5876 &    7.838  \\
48471.9366 &     -163.1004 &    7.653  \\
48471.9419 &     -176.1589 &    6.416  \\
48471.9472 &     -185.5920 &    6.677  \\
48471.9525 &     -194.5284 &    6.225  \\
48471.9578 &     -202.5913 &    6.497  \\
48471.9631 &     -207.7151 &    7.276  \\
48471.9684 &     -213.9204 &    6.869  \\
48471.9737 &     -219.8087 &    6.796  \\
48471.9791 &     -221.0292 &    7.845  \\
48471.9844 &     -227.2314 &    7.481  \\
48471.9897 &     -229.8207 &    7.022  \\
48471.9950 &     -230.1261 &    6.794  \\
48472.0003 &     -232.6394 &    7.163  \\
48472.0079 &     -227.5887 &    6.922  \\
48472.0132 &     -227.9404 &    7.330  \\
48472.0185 &     -222.5029 &    7.479  \\
48472.0239 &     -217.0616 &    7.229  \\
48472.0292 &     -213.9255 &    6.978  \\
48472.0345 &     -207.2891 &    6.660  \\
48472.0398 &     -198.5409 &    6.446  \\
48472.0451 &     -191.5137 &    6.657  \\
48472.0504 &     -182.1613 &    6.712  \\
48472.0557 &     -172.4799 &    7.469  \\
48472.0610 &     -163.6605 &    7.279  \\
48472.0663 &     -153.9607 &    7.165  \\
48472.0718 &     -141.3581 &    7.079  \\
48472.0771 &     -129.7343 &    6.812  \\
48472.0824 &     -114.6732 &    6.248  \\
48472.0896 &      -97.0794 &    6.813  \\
48472.0949 &      -86.5703 &    6.533  \\
48472.1002 &      -72.1837 &    5.909  \\
48472.1055 &      -60.1698 &    6.059  \\
48472.1108 &      -43.6431 &    6.745  \\
48472.1161 &      -31.6927 &    5.927  \\
48472.1214 &      -20.3661 &    7.159  \\
48472.1267 &       -7.2741 &    6.957  \\
48472.1321 &        0.7625 &    6.443  \\
48472.1374 &       14.9686 &    6.286  \\
48472.1427 &       24.2414 &    6.301  \\
48472.1480 &       37.4443 &    5.874  \\
48472.1533 &       42.5435 &    6.033  \\
48472.1586 &       55.0288 &    5.722  \\
48472.1639 &       64.6959 &    5.230  \\
48472.1712 &       74.5404 &    5.386  \\
48472.1765 &       80.6155 &    5.978  \\
48472.1818 &       86.0617 &    5.159  \\
48472.1872 &       90.5772 &    5.348  \\
48472.1925 &       99.3953 &    6.555  \\
48472.1978 &       98.4657 &    6.713  \\
48472.2031 &       98.0316 &    6.387  \\
48472.2084 &       99.6052 &    6.685  \\
48472.2137 &      100.5013 &    6.638  \\
48472.2191 &       99.5221 &   10.041 \\
48472.2244 &       98.9665 &    8.874  \\
48472.2297 &       95.5619 &   10.709 \\
48472.2350 &       86.0348 &   14.289 \\
48472.2403 &       83.6323 &   12.475 \\
48472.2456 &       73.7468 &    8.801  \\
48472.2509 &       67.5799 &    8.262  \\
48472.2563 &       64.1177 &    8.744  \\

            \noalign{\smallskip}
            \hline
         \end{tabular}
      \]
   \end{table}

   \begin{table}
      \caption[]{Radial velocities for AE Aqr for September 22-23, 1997}
         \label{RadVel3}
      \[
         \begin{tabular}{rrr}
            \hline
            \noalign{\smallskip}
        HJD     & Absorption    & $\sigma$   \\
        (240000+)  &    (km\,s$^{-1}$)  \\
            \noalign{\smallskip}
            \hline
            \noalign{\smallskip}
50713.7239 &     53.8834 &   6.056  \\
50713.7434 &     17.1704 &   7.931  \\
50713.7711 &    -57.9052 &  18.501  \\
50713.7850 &    -91.7092 &   8.007  \\
50713.7989 &   -124.5789 &   7.522  \\
50714.6328 &   -160.4805 &   8.160  \\
50714.6446 &   -180.6777 &   7.553  \\
50714.6553 &   -201.3855 &  10.647  \\
50714.6662 &   -213.5401 &   8.437  \\
50714.6853 &   -231.5731 &   8.199  \\
50714.6938 &   -234.5440 &   8.574  \\
50714.7033 &   -238.4612 &   6.789  \\
50714.7122 &   -231.7661 &   7.227  \\
50714.7216 &   -228.0453 &   5.229  \\
50714.7302 &   -221.1735 &   8.284  \\
50714.7389 &   -209.9548 &   7.381  \\
50714.7476 &   -188.2585 &   5.439  \\
50714.8085 &    -49.6652 &  11.475  \\
            \noalign{\smallskip}
            \hline
         \end{tabular}
      \]
   \end{table}

   \begin{table}
      \caption[]{Radial velocities for AE Aqr for August 17-18, 2000}
         \label{RadVel4a}
      \[
         \begin{tabular}{rrr}
            \hline
            \noalign{\smallskip}
        HJD     & Absorption   & $\sigma$  \\
        (240000+)  &     (km\,s$^{-1}$)  \\
            \noalign{\smallskip}
            \hline
            \noalign{\smallskip}
51773.7263 &    65.2917 &    5.101  \\
51773.7363 &    51.2997 &    3.848  \\
51773.7459 &    36.9166 &    4.071  \\
51773.7559 &    15.4658 &    2.995  \\
51773.7655 &    -5.8116 &    3.193  \\
51773.7751 &   -27.7636 &    3.297  \\
51773.7901 &   -70.0045 &    6.230  \\
51773.7997 &   -97.6795 &    6.767  \\
51773.8093 &  -125.9648 &    4.670  \\
51773.8191 &  -148.8826 &    4.394  \\
51773.8287 &  -169.2657 &    4.268  \\
51773.8383 &  -185.6042 &    4.281  \\
51773.8699 &  -223.2806 &    3.845  \\
51773.8794 &  -227.7806 &    4.271  \\
51773.8891 &  -227.2547 &    4.247  \\
51773.9045 &  -224.7867 &    5.136  \\
51773.9141 &  -217.3888 &    4.857  \\
51773.9237 &  -211.7197 &    6.597  \\
51773.9340 &  -191.5169 &    4.304  \\
51773.9436 &  -176.4511 &    4.720  \\
51773.9532 &  -158.7131 &    5.179  \\
51774.6477 &  -159.4111 &    4.662  \\
51774.6669 &  -191.9790 &    4.974  \\
51774.6769 &  -205.5982 &    5.558  \\
51774.6865 &  -217.7922 &    4.667  \\
51774.6961 &  -223.7707 &    4.932  \\
51774.7100 &  -229.8889 &    5.784  \\
51774.7196 &  -226.9689 &    5.189  \\
51774.7292 &  -223.6042 &    4.290  \\
51774.7390 &  -217.1240 &    4.296  \\
51774.7486 &  -207.0125 &    4.437  \\
51774.7582 &  -193.3962 &    4.688  \\
51774.7689 &  -173.2024 &    3.150  \\
51774.7786 &  -159.1386 &    4.645  \\
51774.7881 &  -133.0462 &    3.750  \\
51774.8047 &   -92.2058 &    4.272  \\
51774.8143 &   -65.5297 &    3.668  \\
51774.8239 &   -42.4604 &    3.638  \\
51774.8341 &   -16.3695 &    4.008  \\
51774.8437 &     5.6063 &    3.759  \\
51774.8533 &    25.5118 &    4.780  \\
51774.8636 &    48.4668 &    4.031  \\
51774.8732 &    65.0504 &    4.190  \\
51774.8828 &    76.6211 &    5.118  \\
51774.8925 &    87.8831 &    6.009  \\
51774.9021 &    95.6748 &   13.954  \\
51774.9117 &    98.1195 &   14.311  \\
51774.9261 &    94.3257 &   14.438  \\
51774.9357 &    92.1290 &    7.321  \\
51774.9453 &    83.7146 &   18.339  \\
51774.9559 &    73.6381 &    5.081  \\
51774.9655 &    63.6809 &    7.791  \\
           \noalign{\smallskip}
            \hline
         \end{tabular}
      \]
   \end{table}

   \begin{table}
      \caption[]{Radial velocities for AE Aqr for August 19, 2000}
         \label{RadVel4b}
      \[
         \begin{tabular}{rrr}
            \hline
            \noalign{\smallskip}
        HJD     & Absorption   & $\sigma$  \\
        (240000+)  &     (km\,s$^{-1}$)  \\
            \noalign{\smallskip}
            \hline
            \noalign{\smallskip}
51775.6393 &   -52.7959 &    3.865  \\
51775.6489 &   -28.7739 &    3.291  \\
51775.6585 &    -7.5620 &    4.793  \\
51775.6681 &    13.4982 &    5.374  \\
51775.6999 &    77.3029 &    6.634  \\
51775.7095 &    86.5576 &    7.545  \\
51775.7190 &    99.5335 &   10.032  \\
51775.7353 &   106.2833 &   17.445  \\
51775.7449 &   100.8198 &   17.871  \\
51775.7545 &    98.0898 &   17.983  \\
51775.7650 &    87.9296 &   20.548  \\
51775.7746 &    82.2610 &   15.630   \\
51775.7842 &    66.9225 &   12.160  \\
51775.7979 &    47.5450 &   13.763  \\
51775.8075 &    25.6446 &   12.399  \\
51775.8171 &    11.6025 &   11.819  \\
51775.8268 &   -13.1728 &   14.711  \\
51775.8364 &   -40.5701 &   24.572  \\
51775.8460 &   -62.9008 &   18.127  \\
51775.8581 &   -97.6792 &   23.011  \\
51775.8677 &  -123.7154 &   23.438  \\
51775.8773 &  -144.6751 &   18.331  \\
51775.8914 &  -176.1165 &   17.524  \\
51775.9010 &  -192.9387 &   28.153  \\
51775.9106 &  -209.5299 &   13.797  \\
51775.9210 &  -214.6285 &   21.748  \\
51775.9306 &  -227.1362 &   25.586  \\
51775.9402 &  -229.8586 &   21.955  \\
            \noalign{\smallskip}
            \hline
         \end{tabular}
      \]
   \end{table}

   \begin{table}
      \caption[]{Radial velocities for AE~Aqr for August 29, 2001}
         \label{RadVel5}
      \[
         \begin{tabular}{rrr}
            \hline
            \noalign{\smallskip}
        HJD     & Absorption    & $\sigma$   \\
        (240000+)  &    (km\,s$^{-1}$)  \\
            \noalign{\smallskip}
            \hline
            \noalign{\smallskip}
52150.8019 &    75.3120 &   4.437  \\
52150.8047 &    69.5589 &   5.484  \\
52150.8076 &    63.8477 &   4.453  \\
52150.8104 &    56.6756 &   4.463  \\
52150.8132 &    52.0995 &   4.274  \\
52150.8160 &    49.3786 &   4.825  \\
52150.8188 &    42.0882 &   5.078  \\
52150.8217 &    36.1923 &   5.523  \\
52150.8245 &    33.8993 &   6.534  \\
52150.8273 &    24.1148 &   4.582  \\
52150.8301 &    20.0940 &   5.016   \\
52150.8330 &    11.8558 &   4.700  \\
52150.8358 &    11.2627 &   3.945  \\
52150.8386 &     4.2054 &   4.896  \\
52150.8414 &    -2.7119 &   4.226  \\
52150.8442 &    -9.1846 &   4.445  \\
52150.8471 &   -19.7468 &   4.924  \\
52150.8499 &   -32.0708 &   5.726  \\
52150.8527 &   -34.3096 &   5.178  \\
52150.8556 &   -42.5093 &   5.626  \\
52150.8585 &   -49.2404 &   4.340   \\
52150.8613 &   -59.2369 &   5.035  \\
52150.8641 &   -66.0804 &   5.202  \\
52150.8669 &   -73.8477 &   5.967  \\
52150.8697 &   -84.9643 &   4.653  \\
52150.8725 &   -85.0400 &   5.993  \\
52150.8754 &   -96.7729 &   4.441  \\
52150.8782 &  -104.3313 &   5.963  \\
52150.8810 &  -112.2779 &   4.985  \\
52150.8839 &  -123.8766 &   5.261  \\
52150.8867 &  -130.6497 &   4.880  \\
52150.8895 &  -134.5644 &   4.070  \\
52150.8923 &  -141.7841 &   4.490  \\
52150.8952 &  -150.4175 &   5.517  \\
52150.8980 &  -153.4627 &   5.134  \\
52150.9008 &  -160.6009 &   4.838  \\
52150.9036 &  -164.4972 &   4.363  \\
52150.9064 &  -171.1431 &   6.183  \\
52150.9093 &  -177.4043 &   4.280  \\
52150.9176 &  -190.1421 &   7.178  \\
52150.9204 &  -195.2365 &   5.421  \\
52150.9232 &  -192.0441 &   5.161  \\
52150.9260 &  -204.0816 &   5.099  \\
52150.9288 &  -209.2001 &   8.295  \\
52150.9317 &  -204.1222 &   6.086  \\
52150.9345 &  -213.5170 &   8.890  \\
52150.9373 &  -217.0743 &   9.561  \\
            \noalign{\smallskip}
            \hline
         \end{tabular}
      \]
   \end{table}

To calculate the spectroscopic orbital parameters (assuming a
circular orbit), we have used a least-squares program which
simultaneously fits the four fundamental variables in the equation:
$$V(t)_{\rm abs} = \gamma + K_{\rm abs} \, \sin[(2\pi(t - HJD_{\odot})/P_{\rm orb})],$$
where $V(t)_{\rm abs}$ are the observed radial velocities; $\gamma$ is
the systemic velocity; $K_{\rm abs}$ is the semi-amplitude of the radial
velocity curve; $HJD_{\odot}$ is the heliocentric Julian date of the
inferior conjunction of the companion; and $P_{\rm orb}$ is the orbital
period of the binary.

In Table~\ref{orbB2} we show the orbital parameters
calculated from the SPM and AAT data using 61~Cyg~A and HR~222
standards, respectively. The results for the semi-amplitude are very
similar for both. However, there is a significant difference in the
systemic velocity, probably due to the uncertainties in the absolute
radial velocities of the template stars. We will adopt the systemic
velocity obtained through 61~Cyg~A, since this is a well known primary
standard star (e.g. Keenan \& McNeil 1976; Morgan, Keenan \&
Kellman 1943), which has also been used as a template in
previous studies of AE~Aqr.

\begin{table}
\centering
\begin{minipage}{80mm}
\caption{Orbital Parameters obtained from the absorption lines of
AE~Aqr for SPM and AAT data, using cross correlations with
61~Cyg~A and HR~222 respectively. The sigma values are the
rms errors of the sinusoidal fits.} \label{orbB2}
\begin{tabular}{cccc}
\hline
Orbital          &        AAT            &      SPM             \\
Parameter        &                       &                      \\
\hline
  $\gamma$       & $-69.35 \pm 0.97$     & $-63.50 \pm 0.35$    \\
  (km\,s$^{-1}$) &                       &                      \\
  $K_{\rm abs}$  & $169.74 \pm 1.02$     & $168.72 \pm 0.50$    \\
  (km\,s$^{-1}$) &                       &                      \\
  $HJD_{\odot}$  &  0.951(3)             &  0.78386(4)          \\
  (2439030+)     &                       &                      \\
  $P_{orb}$      &    0.410645(1)        & 0.41165557(3)        \\
   (days)        &                       &                      \\
${\sigma}$       &   6.96                &  3.95                \\
\hline
\end{tabular}
\end{minipage}
\end{table}

\section{Semi-amplitude of the secondary}  \label{rvsec}

\begin{figure}
\vspace{10cm}
\includegraphics[width=1.0\columnwidth]{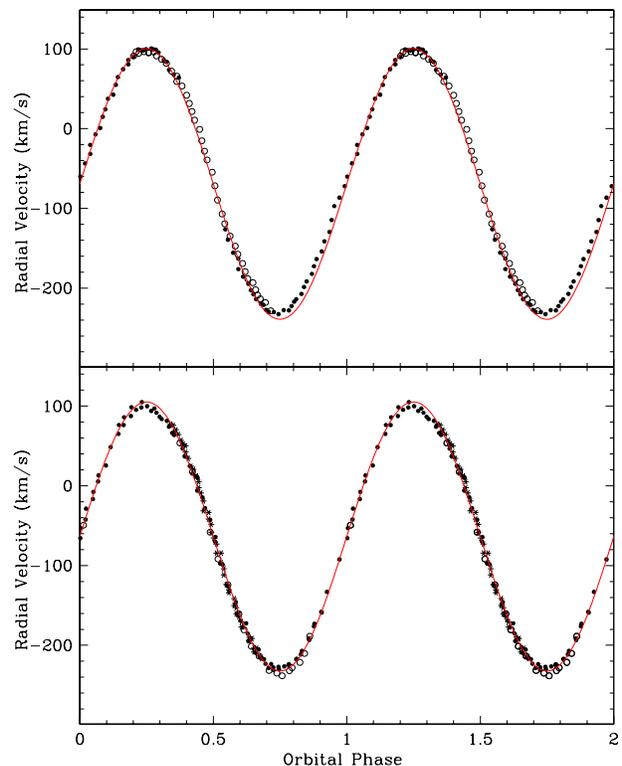}
  \caption{Radial velocity curves of the absorption lines of the secondary
star. The upper plot shows the AAT data, where dots indicate night 1 and open
circles night 2. The lower plot shows the SPM data, where dots
indicate September 1997, open circles indicate August 2000 and
asterisks August 2001. The solid curves are the sinusoidal fits
defined by the parameters in Table 8.}
  \label{fig1}
\end{figure}

Figure~\ref{fig1} shows the radial velocity curves of the secondary,
separately for the AAT (top) and SPM (bottom) data. The orbital
phase was derived from the revised ephemeris in section 6. The
symbols, corresponding to data from different nights, are explained
in the figure caption. The error bars for the individual
measurements are given in the third column of the Radial Velocity
Tables. These are the errors on the centroid obtained by fitting a
Gaussian to the cross-correlation peak in each spectrum. In general
they are of the order of 7\,km\,s$^{-1}$, although there are some
points with errors significantly greater than this mean value. The
sinusoidal (solid) curves have the corresponding semi-amplitude and
systemic values given in Table~\ref{orbB2}.

\begin{figure}
\vspace{10cm}
\includegraphics[width=1.0\columnwidth]{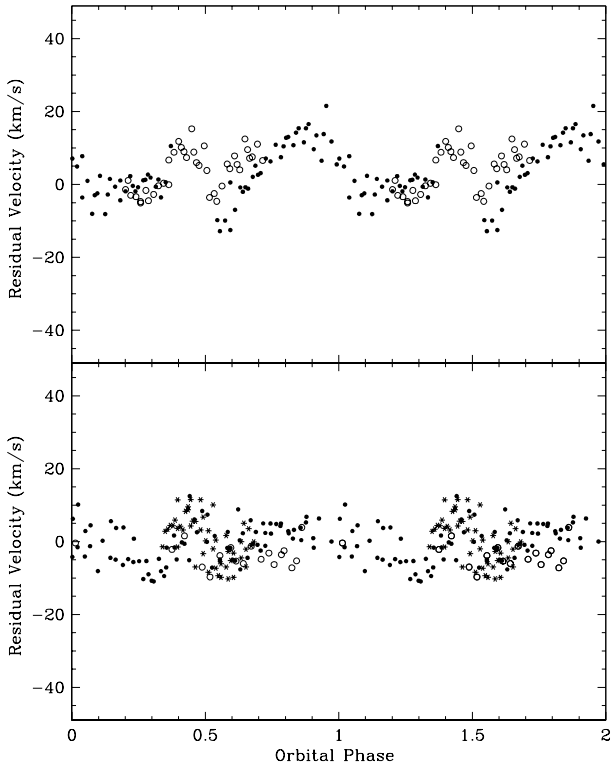}
  \caption{Residual velocities of the absorption lines of the secondary
star. The plotted points are as described in Figure~\ref{fig1}, but here we
have subtracted the fitted sinusoid curves to show the deviations more
clearly.}
  \label{fig2}
\end{figure}

At some orbital phases, the radial velocity points deviate from a
sinusoid. In order to study these deviations more carefully,
we have subtracted the sinusoidal fits described in the last
paragraph and obtained a plot  of the residual velocities, shown in
Figure~\ref{fig2}. For the AAT data (top panel), we observe a fairly
clear double-peaked curve; for the SPM data (lower panel) the curve
is noisier, but there is still a peak apparent at about phase 0.4,
similar to the one in the AAT data. For both sets of data, the peak
is followed by a trough at around phase 0.6. This behaviour is very
reminiscent of the residual velocity curves plotted by Davey \&
Smith (1992); in that paper, for the comparable disk systems with
the best data, IP Peg and YY Dra, there is also a positive peak in
the residuals plot at around phase 0.4 and a negative peak around
phase 0.6. In the Davey \& Smith (1992) paper, this behaviour was
shown to correspond to asymmetric heating of the hemisphere of the
star that faces the disk. That strongly suggests that this feature
of the AE Aqr residual plots also arises from the fact that the
hemisphere facing the disk is being heated. The fact that it occurs
in both data sets suggests that it is a permanent feature of the
system, as one might expect: the radiation source is there all the
time, even if it may vary in strength.

The effect of heating is to reduce the absorption line strength on
the heated inner hemisphere and so to increase the amplitude of the
measured radial velocity curve, because the centre of light moves
towards the averted hemisphere, away from the centre of mass. It is
clear that a cool region of {\it enhanced} absorption line strength
on the unheated outer hemisphere would also increase the amplitude,
for the same reason, but that the effect on the residuals would now
be centred roughly around the position of that region instead of
around $L_1$. To a first approximation then, a region of greater
absorption line strength would produce a second peak and trough
about 0.5 in phase later than the one coming from the heated region
and indeed we see in the AAT data a clear second rise and fall
centred roughly on phase~1. This suggests that at the time of the
AAT observations there was a fairly strong region of enhanced
absorption somewhere on the unheated hemisphere, close to the
`anti-$L_1$ point' in longitude, although we can say nothing about
latitude.  The fact that we do NOT see a similar second peak in the
SPM data would suggest that at the time of the SPM observations
there was no such region (although, since the observations covered a
much larger range of dates, we can't be so confident about that). A
cool, dark feature that is only present on some occasions is
reminiscent of the behaviour of sunspots, and our results may
therefore be consistent with the tomographic observations of a
`star-spot' by Watson et al. (2006). However, our feature would
produce an absorption dip in the absorption lines rather than the
emission bump seen in the tomography data, so the connection is not
obvious.

\section{Ephemeris Revisited}  \label{ephem}

Following the procedure of Welsh et al. (1995) we have extended the
radial velocity data base, starting from the first measurements by
Joy (1954) (105 points over the period 1943 to 1953), and including
the measurements of Chincarini \& Walker (1981) (250 points);
Reinsch \& Beuermann (1994) (40 points); Welsh et al. (1995) (197
points); Casares et al.~(1996)~(207 points); and this paper (247
points, the most recent in 2001). All data were carefully examined,
checking for inconsistencies and typographical errors. In all we
have included 1046 points covering 58 yr of observations. A data
archive of the radial velocity data used here is available on
request to us. We have included in our calculations the published
errors of every measurement. In those cases where they have not been
tabulated, we have taken mean values based on their spectral
resolution and exposure times (40\,km\,s$^{-1}$ for Joy 1954 and
20\,km\,s$^{-1}$ for Chincarini \& Walker 1981). An improved orbital
period of 0.4116554800~d was found. A new zero phase has been
obtained using the SPM data with the new orbital period value. From
these results we adopt the ephemeris

\begin{equation}
\rm{HJD} = 2,439,030.78496(9) + 0.4116554800(2)E .
\end{equation}

\section{Rotational velocities, mass ratio and inclination angle}  \label{rotvel}

\begin{figure}
\vspace{5.5cm}
\includegraphics[width=0.75\columnwidth,angle=270]{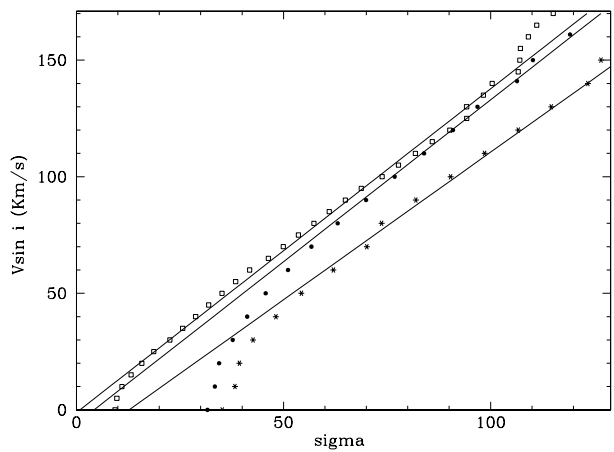}
  \caption{Rotational velocity calibrations derived from the broadening of
template stars. The AAT data are shown as open squares. For the SPM data,
dots denote the Thomson detector and asterisks the SITe detector. The solid
lines correspond to linear fits derived from the data in the 50-100 sigma
range only (see text).}
    \label{fig3}
\end{figure}

Rotational velocities, $V \sin \, i$, of the secondary
star, derived from observations, can lead us to independent
estimates of the mass ratio if $K_2$ is known, and consequently to
the semi-amplitude of the primary star (Horne et al. 1986).
Furthermore, they can be used to help us estimate the
inclination angle of the binary by comparing the variation of $V
\sin \, i$ with orbital phase against models (e.g. Casares et
al. 1996). This approach has been made for AE~Aqr by the latter
authors and also by Welsh et al. (1995), who took a more general
approach in the sense that the amplitude of any ``ellipsoidal
variations'' arising from the secondary can be used to constrain the
inclination angle, because for a given mass ratio, the amplitude of
the variation will depend on the inclination angle of the system.
This will be true not only for rotational velocities, but also for
the continuum variations and the observed absorption-line
flux as a function of orbital phase. In this section we derive $V
\sin \, i$ from our observations and derive probable values for the
mass ratio and inclination angle.

The rotational velocity of the secondary star can be obtained from
the width of the cross-correlation function. We attribute the
increased width of the correlation function with respect to the
intrinsic width of the template star, to the rotational broadening
of the secondary star. To convert the calculated sigma of
the Gaussian fit to the peak of the CCF to $V \sin~i$ we have
broadened the template stars HR222 and 61~Cyg~A with a suitable
rotational kernel.  The kernel was produced for a range of $V \sin
\, i$ from 10 to 200\,km\,s$^{-1}$ and applied to the IRAF program
{\it convolve} to broaden the template star. The broadened
templates are then cross-correlated with the original template and
the $\sigma$ of the Gaussian fit is calculated. We have used a
simple broadening function for spherical bodies as described by Gray
(1976), using a limb darkening coefficient of $\epsilon$ = 0.5 and a
bin width corresponding to the spectral resolution of the AAT
observations and the two spectral resolutions of the SPM spectra
(Thomson and SITe detectors). The resulting calibrations are shown
in Figure \ref{fig3}. The symbols are explained in the figure
caption.

\begin{figure}
\includegraphics[width=1.0\columnwidth]{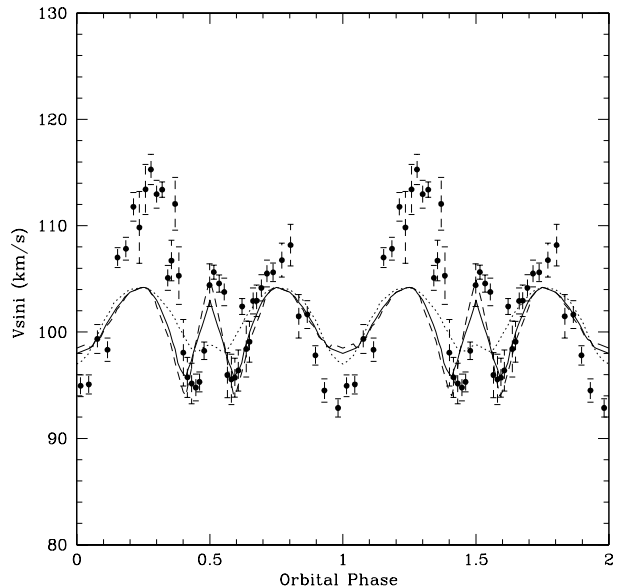}
  \caption{Rotational velocity of the secondary star as a function of orbital phase,
  obtained from the cross correlations derived from the combined AAT and SPM observations.
  The error bars are discussed in the text.}
  \label{fig4}
\end{figure}

We have calculated the rotational velocities $V \sin \, i$ from the
measured sigma of the cross correlations, calibrated with the linear
fits depicted in Figure \ref{fig3}. We believe this is well
justified as the measured sigma values are in the linear region, far
from the non-linear points (shown in Figure~\ref{fig3}) calculated
for lower rotational velocities which, in each case, approach the
velocity resolution for each instrumental setup. At some orbital
phases, the scatter is large, a result also found by Casares et al.
(1996) and by Welsh et al. (1995). As was also done by these
authors, we binned our data to lower the scatter and obtain error
estimates. The rotational velocity results are depicted in
Figure~\ref{fig4}. The error bars are the probable errors,
i.e. the standard deviations for the 5 point samples. These
values may suffer from small number statistics; also the error
estimate assumes that there is no intrinsic variation of $V \sin \,
i$ over 0.02 of an orbit. If there is, we will tend to overestimate
the error bar. Although the AAT and SPM data were combined in the
binning process, and so cannot be distinguished in
Figure~\ref{fig4}, we also looked at the unbinned data and it was
clear that the two datasets were consistent, so that using them
together in the binning was justified. We have produced a mean
rotational velocity value for each of 45 bins, of about 0.02 orbital
phase width each, by calculating the simple average and standard
deviation of a constant sample of 5 data points in each bin. We
discarded the 18 SPM spectra taken in September 1997, which were not
properly calibrated to convert their sigma into rotational
velocities, as well as five other spectra which had a very poor
signal-to-noise ratio.  In the Figure we observe the basic double
modulation expected from an elongated secondary, with low velocities
detected at phases 0.0 and around 0.5, and larger values at phases
0.25 and 0.75. The minimum rotational velocity observed in both
cases is about 92\,km\,s$^{-1}$ at phase 0.0, when the secondary
passes in front of the accretion disc and about the same value near
phase 0.4--0.6, when the secondary is behind the primary star. This
is expected as these are the phases when we should see a more
spherical body. The rotational maxima are asymmetric; the velocities
found for phase 0.25 have velocities near 114\,km\,s$^{-1}$, while
around phase 0.75 they are about 108\,km\,s$^{-1}$. Another
important feature shown in Figure~\ref{fig4} is the behaviour of the
rotational velocity between phases 0.4 and 0.6; an inverted V-shaped
structure is present, with low velocities at phases 0.4 and 0.6 and
a local maximum at phase 0.5. We have also plotted the three
computed models of $V \sin \,i$ by Casares et al. (1996), depicted
in their Figure~5, and compared them with our observations. Some
major points relating to their models are discussed at the end of
this section, and full details can be found in their paper. The
overall observed rotational curves show big differences from the
results of the models. In general the amplitude is much larger than
predicted by the models. Furthermore, the observational data do not
behave symmetrically. As mentioned above, the rotational velocities
are much larger at phase 0.25 than at phase 0.75, and on the other
hand, close to phase 0.0, the observed velocities are much lower
than predicted by the models. One feature that is present in
both models and observations is the inverted V-shaped structure
between phases 0.4 and 0.6. We will also address this point at the
end of this section, when we try to estimate the inclination angle.

We can derive the mass ratio of the binary, and consequently the
semi-amplitude of the radial velocity curve of the white dwarf,
using the rotational velocity of the secondary star. For a
Roche-lobe-filling co-rotating secondary star the predicted
rotational velocity is:
\begin{equation}
V \sin \, i / K_{2} = ( 1 + q) R_2^{RL} /a,
\end{equation}
(Horne et al. 1986). Combining this relation with the analytical
expression by Echevarr{\'\i}a (1983):
\begin{equation}
 R_2^{RL} /a = 0.47459 [q / (1+q)]^{1/3},
\end{equation}
obtained from the tabulations by Kopal (1959), 
we obtain:

\begin{equation}
V \sin \, i = 0.475 K_{2} q^{1/3}(1+q)^{2/3}.
\end{equation}

This equation applies to rotating spheres with equivalent
radii having the same volume as the contact component. The volume
radius is closest to the equatorial radius of the Roche lobe when
viewed at conjunction. Then, to a first approximation we can use the
observed minimum rotational velocity observed at phase 0.0 of
$V({\rm min}) \sin i = 92 \, \pm 3 \, $\,km\,s$^{-1}$, and use the
equation above. Taking $K_{R} = 168.7 $\,km\,s$^{-1}$ we obtain
$q=0.6 \, \pm 0.02$. We have taken the same approach as Welsh et al.
(1995), but using a slightly different $R_2^{RL} /a$ approximation.

A mass ratio $q=0.6$ implies $K_1 = 101 \pm 3$\,km\,s$^{-1}$,
a value which is in excellent agrement with the $K_1 = 102 \pm
2$\,km\,s$^{-1}$ determination from the time-delay curve of the 33\,s
pulsation reported by Eracleous {\it el al.} (1994).

The inclination angle can be estimated from the amplitude of the
rotational velocity curve. Although the curve is highly distorted
from the ellipsoidal, or tear-drop, shape of the Roche-Lobe, we can
use the maximum and minimum $V \, \sin \, i$ at phases 0.75 and 0.0
if we assume they are representative of the ellipsoidal variations
(108 and 92\,km\,s$^{-1}$ respectively). Taking these values, we
obtain a total amplitude of about 20 percent. We can compare this
result with the ellipsoidal variation models by Welsh et al. (1995)
which are absorption-line flux models constructed for a limb- and
gravity-darkening Roche-Lobe star, using $\beta =0.08$ for a
gravity-darkening given by $T_{\rm{eff}} \, \alpha \, g^{\beta}$.
Their models used $q=0.65$ and a limb-darkening coefficient of
$\epsilon$ = 0.4, but they found that nearly identical estimates
were obtained for $\epsilon$ = 0.6. Their model curves are shown in
the lower part of their Figure 7 for three inclination angles:
$35^\circ$, $50^\circ$ and $65^\circ$. The latter solution has an
amplitude of about 20 percent in agreement with our observations.
Although different observables are being considered, we expect a
similar amplitude, because in both cases the maximum (minimum)
should occur when the largest (smallest) area is seen, and it is the
variation in projected stellar radius that provides the dominant
effect. The amplitude is consistent with the ratio of the mean
radius to the front-back equivalent Roche-Lobe radius for $q\approx
0.6$ (see Figure 4 in Welsh et al. 1995). Further, an observed
amplitude of 20 percent is consistent with the results found by van
Paradijs et al. (1989) both from models and from the observed
amplitude of the quiescent $V$-band ellipsoidal variations. Similar
results are found in the papers by Chincarini \& Walker (1981) and
by Bruch (1991). An interesting feature in all these publications is
that they show that the ellipsoidal photometric variations are not
symmetric; typically, the maximum at phase 0.25 is stronger than the
one at phase 0.75 (see for example Figure 4 in van Paradijs,
Kraakman H. \& van Amerongen 1989), similar to what we find in the
rotational velocity curves.

Is there a possible explanation for the asymmetry? There seem to
be two possible explanations for a larger amplitude for $V \sin i$
at phase 0.25. One is that there is some additional source of
broadening that is stronger on the trailing hemisphere, and the other
is that slightly asymmetric irradiation causes the effective radius of
the star to be larger on the trailing hemisphere than on the leading one.
We shall discuss these possibilities further in Section~12.

Another estimate of the inclination angle can be made by comparing
our observations with the models of Casares et al. (1996).
Their models of $V \sin i$ as a function of phase assume a
quadratic limb-darkening law and a gravity-darkening law given by
$T_{\rm{eff}} \, \alpha \, g^{\beta}$. They computed 16 sets of
models for inclinations and $\beta$ in the range
$40^\circ$--$70^\circ$  and 0--0.3 respectively. The other
parameters were fixed, with $\epsilon$ = 0.65, $T_{\rm{eff}} = 4500$
K, $K_{2} = 162 $\,km\,s$^{-1}$ and $q=0.62$. Their three plotted
models (see their Figure 5) are all for $\beta =0.08$. They find
that the synthetic $V_{rot} \, \sin \,i$ curves exhibit gradual
changes in shape and amplitude as the inclination angle increases.
Their models are shown in Figure~\ref{fig4} for different
inclination angles (dotted line, $40^\circ$; solid line $58^\circ$;
and dashed line $70^\circ$). Although the amplitude of the observed
variation of $V_{rot} \, \sin \, i$ as a function of phase is
greater than those predicted from their models, it is evident that
the V-shaped structure increases in strength with the inclination
angle. Taking this into account, we believe that an
inclination close to $70^\circ$ yields a better fit to our data
than the other values. The more elaborate Roche Tomography
of Watson et al. (2006), which effectively takes two of
the pieces of information we have been using, the variation in line
velocity and line shape/width, and adds a third, the variation in
line flux, which we do not have at our disposal in this paper,
allows them to construct surface maps of the secondary star, from
which they determine the most accurate system parameters. In
particular they calculate an inclination angle of $66^\circ$, close
to our result. Indeed, these authors refer to an unpublished version
of this paper, to support their result on the inclination angle. It
is satisfactory that the different methods used to obtain the
inclination of this binary are now in generally good agreement. We
note that the preferred value is now very close to the upper limit
of $70^\circ$ imposed by the lack of eclipses (Robinson et al.
1991).

\section{Spectral type as a function of orbital phase}  \label{sptype}

\begin{figure}
\vspace{5.6cm}
\includegraphics[width=0.78\columnwidth,angle=270]{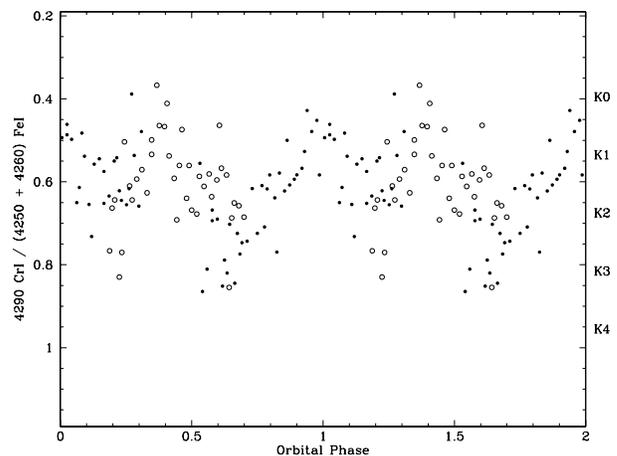}
  \caption{The absorption line ratio Fe~I/Cr~I as a function of orbital phase for the AAT
  observations. This line ratio is a basic temperature indicator for K stars (see text). A first
  order calibration of the spectral type, using the spectral template stars, is shown at the right.}
  \label{fig5}
\end{figure}

The AAT spectra have the highest spectral resolution and best
signal-to-noise ratio for measuring the line depths of the
Fe~I $\lambda\lambda 4250,4260$  and Cr~I $\lambda4290$
absorption lines. These lines are amongst those recommended
for spectral classification in K stars by Keenan \&\ McNeil (1976)
because the atoms belong to the same abundance group and so the line
ratios define spectral type (or effective temperature) essentially
independently of metallicity. We have done this both for the
individual AE~Aqr spectra and for the comparison stars.
Figure~\ref{fig5} shows the line ratio
Fe~I$\lambda\lambda4250,4260$/Cr~I$\lambda4290$ as a
function of orbital phase. A first order calibration of this line
ratio with the same line ratio measured in the spectral standards
(see Table~1) was made. This is shown on the right hand side of the
diagram. We observe that the line ratio has a roughly double
sinusoidal behaviour with two minima and two maxima. It is important
to note that the system appears to have an earlier spectral type
(about K0), close to phases 0.4 and 0.9 and a tendency to
move to later spectral types towards phase 0.2 (about K2) and phase
0.65 (close to K4). These changes in absorption line strengths may
be due to irradiation effects (see Warner 1995 and references
therein), or star-spots on the secondary star (Watson et al. 2006).
These effects as well as the changes in temperature
as a function of phase will be discussed in section~12.

\section{The Fundamental Parameters of AE Aqr}  \label{param}

Assuming that the radial velocity semi-amplitudes reflect accurately
the motion of the binary components, then from our results: $K_{em}
= K_1 = 101 \pm 3$\,km\,s$^{-1}$; $K_{abs} = K_2 = 168.7 \pm
1$\,km\,s$^{-1}$, and adopting $P=0.4116554800$\,d we obtain:

\begin{equation}
M_1 \sin^3 i = {P K_2 (K_1 + K_2)^2 \over 2 \pi G} = 0.52 \pm 0.03
M_{\odot},
\end{equation}

\begin{equation}
M_2 \sin^3 i = {P K_1 (K_1 + K_2)^2 \over 2 \pi G} = 0.31 \pm 0.02
M_{\odot},
\end{equation}
and

\begin{equation}
a \sin i = {P (K_1 + K_2) \over 2 \pi} = 2.19 \pm 0.02 R_{\odot}.
\end{equation}
Assuming an inclination angle $i =70^\circ \pm 3^\circ$, the system
parameters become: $ M_1 = 0.63 \pm 0.05 \, M_{\odot}$; $ M_2 = 0.37
\pm 0.04 \, M_{\odot}$; and $ a = 2.33 \pm 0.02 \, R_{\odot}$.

\section{The $M_1 - M_2$ Diagram}  \label{m1m2}

\begin{figure}
\vspace{8cm}
\includegraphics[width=1.0\columnwidth]{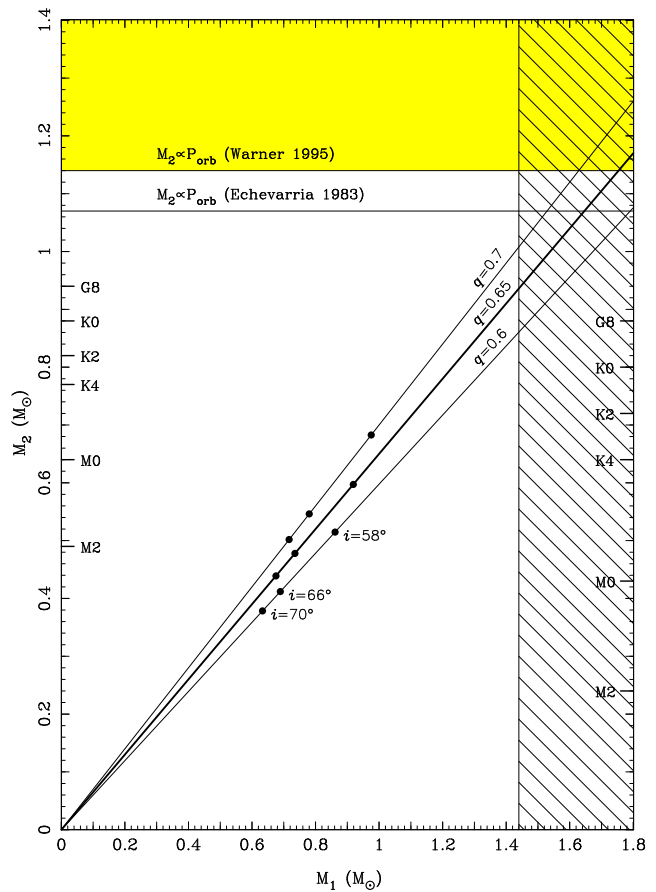}
  \caption{$M_1 - M_2$ mass diagram for AE~Aqr. The components must lie in the
  lower left quadrant of the diagram. The spectral type of the secondary would
  be given by the labels on the left-hand axis if it were a ZAMS star, or by
  the labels on the right-hand axis if an evolved MS star; see text for details.}
  \label{fig6}
\end{figure}

A convenient tool for the analysis of the masses and inclination
angle of the binary has been developed by Echevarr\'{\i}a et al.
(2007). This $M_1$ -- $M_2$ diagnostic diagram is shown in
Figure~\ref{fig6}. In this diagram, the slope of the straight lines
is simply given by $q$ and therefore, for a given value of $M_1$, we
have a corresponding value of $M_2$. The vertical dashed area
indicates a neutron star regime, set at the left by the white dwarf
Chandrasekhar limit. The horizontal upper region shows the
values obtained from the empirical $M_2$--$P_{orb}$ relations by
Warner (1995, page 111) and Echevarr\'{\i}a (1983). These two
relations, using the improved orbital period, give $M_2=1.14 \,
M_{\odot}$ and $M_2=1.07 \, M_{\odot}$ respectively. The spectral
types labelled on the left side of the diagram correspond to the
ZAMS stellar models discussed by Kolb \& Baraffe (2000), while those
shown on the right side correspond to evolved stars that are
still within the main sequence band (see Kolb \& Baraffe 2000 for
detailed explanations). Points in the white area are then within
these limits as we expect that the primary star will not exceed the
Chandrasekhar limit, and that the secondary star will not be on the
MS (Echevarr{\'\i}a 1983; Beuermann et al. 1998). From the three
selected mass ratios ($q$~=~0.7, 0.65 and 0.6), and possible
inclination angles ($i=58^\circ, 66^\circ$ and $70^\circ$), values
which are consistent with most of the published observations shown
in Table~\ref{fundpar}, we conclude that the white dwarf is probably
not a very massive primary, with an upper limit of about $M_1=1.0 \,
M_{\odot}$ (for $i=58^\circ$) and a lower limit of $M_1=0.63 \,
M_{\odot}$ (for $i=70^\circ$). The observational evidence for a high
inclination angle from Robinson et al. (1991), from Watson et al. (2006)
and from this paper suggests that we may be closer
to the lower mass limit. This is also consistent with the expected
lower limit of $0.6 \, M_{\odot}$ for a Helium White Dwarf, in order
to support the 33~s pulsations (Robinson et al. 1991).
Moreover, the observed spectral type is inconsistent with
the predicted values in the diagram, except for the case of a high
mass ratio and low inclination angle solution compared with evolved
main-sequence stars. Furthermore, the possible values for the mass
of the secondary are very different from those of a main sequence
star, as they are far from the upper gray zone in the diagram. This
indicates that we are dealing with a late-type star which is
under-massive for its size or (more plausibly) over-sized for its
mass. Although this conclusion is not new, and has been pointed out
as early as Crawford \& Kraft (1956); Patterson (1979) and
Echevarr{\'\i}a (1983), and later by other authors (see references
in Table~\ref{fundpar} and further discussion in the next section),
it is presented here in the formal context of the $M_1$ -- $M_2$
diagnostic diagram.

\section{The radius of the secondary}  \label{secrad}

Following our results we can compare the mass and radius of the
secondary with those of normal main sequence stars. Using the
mass-radius relation for main sequence stars:
\begin{equation}
{R / R_{\odot}} = 1.057 ({M / M_{\odot}})^{0.906}
\end{equation}
by Echevarr\'{\i}a (1983), and taking the derived mass of the
secondary $M_2=0.37 \, M_{\odot}$ in this paper, we obtain a radius
$R_2^{ms}=0.43 \, R_{\odot}$. On the other hand, using eq.(3) we
obtain: $R_2^{RL}/a = 0.34$, and taking our result on the separation
of the binary $ a = 2.33 \, R_{\odot}$, we obtain $R_2^{RL}=0.79 \,
R_{\odot}$. Taking the ratio $f=R_2^{RL}/R_2^{ms}$ we obtain a value
of 1.84. We will call this ratio the filling factor. Therefore the
secondary star appears to have a radius larger than a star of that
mass on the main sequence by nearly a factor of two. We have
followed the same procedure for all entries in Table~\ref{fundpar}
(see next section), calculating first $R_2^{ms}$ from eq.(8) and the
binary separation $a$ from eq. (7), while $R_2^{RL}$ has been
derived from eq.(3), taking, in both calculations, the corresponding
values in Table~\ref{fundpar}. The results are also shown in
Table~\ref{fundpar}.

\begin{table*}
\centering
\begin{minipage}{180mm}
\caption{Comparison of Fundamental Parameters of AE~Aqr}
\label{fundpar}
\begin{tabular}{@{}lllccccllllrll@{}}
\hline
$P_{orb}$       & $K_2$    & $K_1$       & $q$  & $i$   & $a$  & $V_{rot}\sin \,i$ & $M_2$ & $M_1$ & $R_{2}^{RL}$ & $R_2^{ms}$ & $f$ & Sp.  & Authors \\
(days) & \multicolumn{2}{c}{(km\,s$^{-1}$)}&  & ($^\circ$) & ($R_{\odot}$)& (km\,s$^{-1}$) & ($M_{\odot}$) & ($M_{\odot}$) &  \multicolumn{2}{c}{($R_{\odot}$)} & &  Type  & \\
\hline
0.7007          & 146        &   151        & 1.03 &        &      &        &         &       &      &      &       &  dK0     & J54        \\
0.4118$^a$      & 162.5      &   145        & 0.89 &        &      &        &         &       &      &      &       &          & PG69       \\
0.4116550       & 162.5$^b$  &   129$^*$    & 0.79 & 58     & 2.80 &        &  0.74   & 0.94  & 1.01 & 0.80 & 1.27  &  K5V     & P79        \\
0.4116537       & 159        &   135        & 0.85 & 64     & 2.66 &        &  0.69   & 0.82  & 0.98 & 0.75 & 1.31  &  K5V     & CW81       \\
0.4116580       & 160$^c$    &   141        & 0.88 & 63$^-$ & 2.75 &        &  0.76   & 0.88  & 1.01 & 0.82 & 1.23  &  K5V$^h$ & RSB91      \\
0.4116580       & 160$^c$    &   141        & 0.88 & 70$^+$ & 2.60 &        &  0.65   & 0.75  & 0.96 & 0.71 & 1.35  &  K5V$^h$ & RSB91      \\
0.41165561      & 159        &   122$^d$    & 0.78 & 57     & 2.72 &        &  0.70   & 0.91  & 0.98 & 0.76 & 1.29  &  K3V     & RB94       \\
0.411655601$^c$ & 157.9$^e$  &   102$^f$    & 0.65 & 55     & 2.58 & 85-108 &  0.57   & 0.89  & 0.90 & 0.63 & 1.43  & K3-K5    & WHG95      \\
0.411655653     & 162.0      &   102$^f$    & 0.63 & 58     & 2.53 &  101   &  0.50   & 0.79  & 0.88 & 0.56 & 1.57  &  K4      & CMMH96     \\
0.411655653     & \textbf{\it 167.6$^{**}$}      &   \textbf{\it 113.3$^{**}$} & 0.68 & 66     & \textbf{\it 2.50$^{**}$} &  \textbf{\it 99$^{**}$}   &  0.50   & 0.74  & {\it 0.88$^{**}$} & {\it 0.56$^{**}$} & {\it 1.57$^{**}$}  &  K4      & WDS06      \\
0.4116554800    & 168.7      &   101$^g$    & 0.60 & 70     & 2.33 &   92   &  0.37   & 0.63  & 0.79 & 0.43 & 1.84  & K0-K4    & This paper \\
\hline
\end{tabular}
Adopted from a) Walker (1965); b) Payne-Gaposchkin (1969); c) from
Feldt \& Chincarini (1980); d) from $K_{pulse}$ by Robinson et al.
(1991); e) from Welsh et al. (1993); f) from $K_{pulse}$ by
Eracleous et al. (1994); g) derived from our measured rotational
velocity and deduced mass ratio (see Section \ref{rotvel}); h)
Chincarini \& Walker (1981); *) pulse arrival method. **) Inferred
from their entropy landscape results -- see text for details. -)
minimum and +) maximum inclination angle limits.
\\References: J54: Joy (1954); PG69: Payne-Gaposchkin (1969); P79:
Patterson (1979); CW81: Chincarini \& Walker (1981); RSB91: Robinson
et al. (1991); RB94: Reinsch \& Beuermann (1994); WHG95: Welsh et
al. (1995); CMMH96: Casares et al. (1996); WDS06: Watson et al.
(2006). The results by Watson et al. (2006) are the only ones that
make a correction for the effects of variable intensity over the
surface of the secondary.

\end{minipage}
\end{table*}

\section{Discussion}  \label{discus}

Table~\ref{fundpar} shows the fundamental parameters of the binary,
obtained by several authors since the early work by Joy (1954).
Columns (1) to (14) show: the period; the measured (and assumed)
semi-amplitudes $K_2$ and $K_1$ for the secondary and primary stars
respectively; the mass ratio $q$ and the estimated inclination angle
$i$; the binary separation $a$ (which has been calculated here when
necessary); the projected rotational velocity of the secondary; the
masses of the secondary and primary components; the Roche-Lobe
radius $R_2^{RL}$ and the corresponding main-sequence radius, based
on the estimated mass of the secondary; the filling factor $f$; the
estimated spectral type and the references to the authors. We can
observe in the Table a certain trend with time. Since the first
observations by Joy (1954), there is a slight tendency towards
larger radial velocity semi-amplitudes for the secondary star, a
stronger tendency for lower values for the primary star, and hence a
steady decrease in mass ratio. The secondary radii and calculated
filling factors depend strongly on the mass of the primary and on
the mass ratio, and therefore there is also a decrease in the
main-sequence radius and an increase in the filling factor. These
tendencies cannot be attributed to real changes with time, but
rather to a refinement in the measured parameters as a result of
better observing techniques and the use of instruments with higher
spectral resolutions. The latest results, by Watson et al. (2006)
and this paper, differ mainly because of the adoption
of different values for the semi-amplitude of the primary star.
These estimates are derived from two new techniques: the entropy
landscape and the rotational velocity of the secondary respectively.
Although entropy maps do in principle give the most ``accurate''
values of orbital parameters, their errors are very difficult to
assess properly, because enormously time-consuming Monte-Carlo
calculations would be needed (Watson et al. 2006; see,
however, Watson et al. 2007 for a way of estimating the error). On
the other hand, our result for $K_{wd}$ depends on an educated guess
as to which is the rotational velocity that gives the least-deformed
secondary, among the values of this complex and variable parameter
as a function of orbital phase. However, our result is in good
agreement with the results by Eracleous et al. (1994), based on the
independent $K_{\rm pulse}$ estimate. We have obtained also a very
large filling factor, greater than previously estimated. Such an
increase in the radius of the secondary star could be explained
through X-ray heating mechanisms discussed by Hameury et al. (1993),
while evolved main sequence star scenarios will also yield a
significantly larger stellar radius (Kolb \& Baraffe 2000); in
addition, any mass-transferring star is likely to be out of thermal
equilibrium, increasing the radius as a result.

The paper by Watson et al. (2006) only quotes explicitly the values
for the masses and inclination that the authors deduce from their
entropy landscape technique. This technique uses the currently most
sophisticated procedure to allow for the variation in intensity over
the surface of the secondary, and their results are probably the
most reliable ones in Table \ref{fundpar}. To enable us to compare
their results with our own, and with previously published results,
we have used their masses and inclination, together with equations
(3) to (8) in the present paper, to compute the other entries in
their row of Table 9; these entries are in italics to emphasise that
they have been derived by us and are not quoted in their paper (the
spectral type of K4 is assumed in their paper, so is not in
italics). The most noticeable difference is that the $K_1$ value
deduced from their paper is significantly larger than ours and than
the $K_1$ deduced from spin pulse measurements (Eracleous et al.
1994).

As discussed in Section 5, the observed radial velocity residuals
seem to be explicable by a combination of irradiation effects on the
hemisphere of the secondary facing the white dwarf and some form of
enhanced absorption on the averted hemisphere.

A full discussion on whether the radial velocity semi-amplitude of
the secondary star in a cataclysmic variable increases with
irradiation or back-illumination effects, or decreases with line
quenching, has been made by several authors (e.g. Wade 1981; Martin
1988; Friend et al. 1990). However, in the case of AE~Aqr an
estimation of this effect is not easy to obtain. We found that, from
the published results shown in Table~\ref{fundpar}, only Watson et
al. (2006) make a correction, based on the entropy landscape
technique. Although they do not give an explicit value for the
corrected $K_2$, it is possible, as discussed above and shown in
Table~\ref{fundpar}, to derive it from their given masses and
inclination angle. Since they quote an uncorrected $K_2 = 168.4 \pm
0.2 $\,km\,s$^{-1}$ value, we estimate that their correction amounts
to 0.8 \,km\,s$^{-1}$ only. We are unable to make a proper
correction to our data, due to the complex behaviour of the velocity
residuals shown in Figure~\ref{fig2} and discussed in Section
\ref{rvsec}. Whether or not the underlying sinusoid in
Figure~\ref{fig1} may be a reasonable representation of the orbital
motion or not, nevertheless, our uncorrected $K_2$ result does not
differ greatly from the corrected value by Watson et al. (2006).
Here, we may safely assume that our $K_1$ result, based on the mass
ratio deduced from the rotational velocities and the semi-amplitude
of the secondary, will not be significantly affected. We can indeed
use their ``corrected'' $K_2$ value in equation (4) and obtain
essentially the same result for $K_1$ as we find from our
uncorrected value of $K_2$.

Irradiation effects near the inner Lagrangian point were also found
by Watson et al. (2006), and they could partially explain the
complex temperature behaviour as a function of orbital phase. At
phase 0.0 we would observe an unirradiated hemisphere of the
secondary with a temperature equivalent to a K0 star. However, where
irradiation effects are important, there may be a quenching effect
on the Fe~I absorption lines greater than on the Cr~I lines,
producing (paradoxically) an apparent decrease in temperature and
moving the spectral type to later classes on the irradiated
hemisphere. The source of the enhanced absorption is harder to
understand. Although Watson et al. (2006) found strong evidence for
a spotted region at high latitude on the secondary, at a longitude
that corresponds to its being on the meridian roughly at orbital
phase 0.25, there is no sign in their Roche tomograms of any
enhanced absorption on the averted hemisphere. The source of the
enhanced absorption that seems to be required by the velocity
residuals thus remains uncertain, but is probably not a starspot, at
least in the sense used by Watson et al. (2006).

Can we relate the variations of $V \sin i$ with orbital phase, shown
in Figure \ref{fig4}, to the variations in the radial velocity, and
gain any further information about the region of enhanced
absorption? Certainly, the amplitude of the variations, especially
around phase 0.25, is larger than expected from straightforward
models of the effect of varying projected radius. In other words,
around orbital phases 0.25 and 0.75 the absorption lines appear to
be broader than expected. Irradiation can perhaps account for the
{\it asymmetry} between these phases, for example if the leading
hemisphere were hotter than the trailing one (as found for several
CVs by Davey \& Smith 1992; see also Smith 1995) then the absorption
lines would be quenched preferentially on the leading hemisphere,
near the L$_1$ point, and the line would appear somewhat narrower
(smaller effective $V \sin i$) at phase 0.75 than on the trailing
hemisphere at phase 0.25. However, it is hard to see how irradiation
could ever cause the absorption lines to appear broader than
expected from a purely geometrical effect. The larger-than-expected
amplitude of the curve in Figure~\ref{fig4} seems to require some
additional source of broadening, or some intensification of the line
that has the same effect when it comes to measuring $V \sin i$. A
region of enhanced absorption on the averted hemisphere, centred
some 180$^\circ$ in longitude away from the L$_1$ point, might
provide this, as well as accounting for the velocity residuals in
Figure~\ref{fig2}. However, this sheds no additional light on the
source of the enhanced absorption. The other possible explanation
for the large amplitude is that there is some additional physical
mechanism operating that adds to the width of the line, such as
turbulence, but it would need to be anisotropic, so that the
broadening effect was enhanced at the quadratures; there is no
obvious reason why that should be so.

We conclude that our observations are consistent with, and indeed seem
to require, a region of enhanced absorption on the averted hemisphere
of the secondary star, but that they provide no convincing evidence for
the source of that absorption.

\section{Conclusions}  \label{concl}

High-dispersion time-resolved spectroscopy of the cataclysmic
variable AE~Aqr was obtained during several observing runs over a
ten-year period. The high spectral resolution coupled with the high
signal-to-noise ratio obtained for the strong absorption lines
allows us to find several fundamental parameters of the binary. The
radial velocity analysis yields a semi-amplitude value of $K_{abs} =
168.7 \pm 0.5$\,km\,s$^{-1}$ for the secondary star and a systemic
velocity of $-63$\,km\,s$^{-1}$. A new ephemeris was calculated by
expanding the 39-y radial velocity database of Welsh et al. (1995)
to 58\,y with new observations by Reinsch \& Beuermann (1994),
Casares et al. (1996) and this paper. We have measured the
absorption strengths and line ratio of
Fe~I$\lambda\lambda4250,4260$/Cr~I$\lambda4290$ and made a first
order calibration with spectral type, using the spectral template
stars. The line ratio varies as a function of orbital phase,
equivalent to a variation from K0 to K4 with a mean value of K2. The
rotational velocity of the red star has been measured as a function
of orbital period and shows ellipsoidal variations at twice the
orbital frequency. Using the models by Casares et al. (1996) we
estimate an inclination angle close to $i = 70^\circ $. The
rotational velocities are used to constrain the system mass ratio
and yield a white dwarf semi-amplitude value of $K_{em} =
101$\,km\,s$^{-1}$ consistent with the derived value from the
spin-pulse results by Eracleous et al. (1994).  From these values we
estimate the masses of the binary as $ M_1 = 0.63 \pm 0.05 \,
M_{\odot}$; $ M_2 = 0.37 \pm 0.04 \, M_{\odot}$; and a separation of
$ a = 2.34 \pm 0.02 \, R_{\odot}$. An analysis using the $M_1-M_2$
diagram points towards a secondary star which is over-sized for its
mass, with a radius greater than that of a normal main sequence star
by a factor of almost two. Finally, we discuss the measured
temperature and spectral variations as a function of orbital phase
and suggest that these may be the result of changes in the observed
line ratios due to the presence of regions of enhanced irradiation,
near the inner Lagrangian point but perhaps preferentially on the
leading hemisphere of the secondary, and of enhanced absorption on
the averted hemisphere. The asymmetry in the variation of the
rotational velocity with orbital phase may also be related to the
presence of these regions. We currently have no physical model for
the region of enhanced absorption, but it needs to produce
absorption lines that are abnormally deep or broad, or both.

\section*{Acknowledgments}
We are grateful for the use of the facilities at the Anglo-Australian
Telescope on Siding Spring mountain, operated by the Anglo-Australian
Observatory, and at the 2.1-m telescope at San Pedro M\'artir, operated
by the Mexican Observatorio Astron\'omico Nacional. We acknowledge the use
of software developed by the UK Starlink project and by the NOAO. We thank
the referee for very careful and detailed comments that have led to
significant improvements in the paper.

\bsp

\label{lastpage}

\end{document}